\documentclass{iopart}
\usepackage{graphicx}
\usepackage{amssymb}
\newcommand{\bra}[1]{\langle#1|}
\newcommand{\ket}[1]{|#1\rangle}

\newcommand{\out}[2]{\ket{#1}\bra{#2}}
\newcommand{\expect}[1]{\langle#1\rangle}
\newcommand{\mi}{\mathrm{i}}
\newcommand{\alphav}{\mbox{\boldmath$\alpha$}}
\newcommand{\betav}{\mbox{\boldmath$\beta$}}
\newcommand{\Gv}{\mbox{\boldmath$G$}}
\newcommand{\Iv}{\mbox{\boldmath$I$}}
\newcommand{\Av}{\mbox{\boldmath$A$}}
\newcommand{\Bv}{\mbox{\boldmath$B$}}
\newcommand{\Cv}{\mbox{\boldmath$C$}}
\newcommand{\Sv}{\mbox{\boldmath$S$}}
\newcommand{\Xv}{\mbox{\boldmath$X$}}
\newcommand{\Uv}{\mbox{\boldmath$U$}}
\newcommand{\Zv}{\mbox{\boldmath$Z$}}
\newcommand{\rv}{\mbox{\boldmath$\rho$}}
\begin{document}
\title{Modematching an optical quantum memory}
\author{J Nunn$^1$, I A Walmsley$^1$, M G Raymer$^2$, K Surmacz$^1$, F C Waldermann$^1$, Z Wang$^1$ and D Jaksch$^1$}
\address{$^1$Clarendon Laboratory, Oxford University, UK}
\address{$^2$Oregon Center for Optics and Department of Physics
University of Oregon, Eugene OR 97403, USA}
\ead{j.nunn1@physics.ox.ac.uk}
\begin{abstract}
We analyse the off-resonant Raman interaction of a single broadband photon, copropagating with a classical
`control' pulse, with
 an atomic ensemble. It is shown that the classical electrodynamical structure of the interaction guarantees canonical evolution of the quantum mechanical field operators. This allows
the interaction to be decomposed as a beamsplitter transformation between optical and material excitations on a
mode-by-mode basis. A single, dominant modefunction describes the dynamics for arbitrary control pulse shapes.
 Complete transfer of the quantum state of the incident photon to a collective dark
state within the ensemble can be achieved by shaping the control pulse so as to match the dominant mode to the
temporal mode of the photon. Readout of the material excitation, back to the optical field, is considered in the context of the symmetry connecting the input and output modes. Finally, we show that the transverse spatial structure of the interaction is characterised by the same mode decomposition.
\end{abstract}
 %\submitto{\NJP} 
\maketitle
\section{Introduction}
 Distributed quantum computing \cite{distributed} and quantum cryptographic protocols
\cite{cryptography} require the transmission of quantum
 information, or entanglement, over large distances. The natural candidate for a transmission line of this kind
 is optical fibre, with qubits encoded in the states of broadband single-photon wavepackets. The ability to transfer such `flying' qubits to a material system, where they could be stored or
 manipulated in a controlled way, would greatly facilitate the implementation of many powerful protocols
 \cite{repeater}\cite{grover1}\cite{shor1} for quantum information processing.

\begin{figure}[h]
   \begin{center}
   \includegraphics[width=10cm]{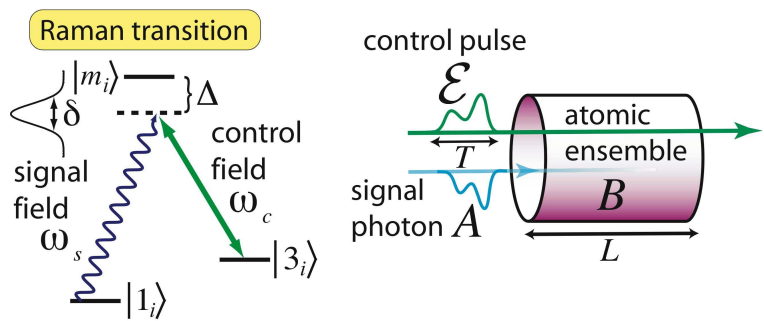}
   \caption{Left: the level structure of the atoms comprising a quantum memory for broadband photons, with bandwidth $\delta$. Right: a schematic
    of the readin process for the quantum memory.}
   \label{Figure1}
   \end{center}
   \end{figure}

   Here we present a scheme for a quantum memory, in which an ultrashort single-photon \emph{signal} pulse is
    coherently absorbed within an atomic vapour through a two-photon Raman transition (see figure
       \ref{Figure1}). This idea has been explored
     previously for the case of narrowband photons \cite{polzikscheme}, drawing on the dynamical theory developed by Raymer and Mostowski \cite{Raymer1}. By analysing the problem in terms of its fundamental mode structure, we show that the main difficulty with this previous proposal arose from poor modematching, because of the use of narrowband fields. More recently, this theory was revisited \cite{raymer2}\cite{Wojtek} in an extension of the entanglement generation protocol proposed by Duan, Cirac, Lukin and Zoller \cite{longdistance}\cite{3Dpaper}. In this paper we investigate a related but distinct dynamics, which affords the ability to store, rather than generate, an entangled photon. In this sense our scheme is similar to that of Fleischhauer et al. \cite{EIT1}\cite{EITreview2}\cite{gorshkov}, based on Electromagnetically Induced Transparency. However, in our scheme there is no group velocity reduction for the signal photon. This implies that the scheme is best suited to short photon wavepackets whose length in space is less than that of the atomic medium. The storage does not rely on the resonant phenomenon of quantum interference, and this allows for the mapping of temporally short, broadband wavepackets into the memory.
     In our proposal the spatio-temporal structure of the signal photon is transferred by a strong classical \emph{control} pulse
     to a long-lived collective
     atomic excitation \cite{EIT1}, which we call a \emph{spin wave}. This is consistent with the fact that the internal
      atomic degree of freedom is often, but not always, an angular momentum. After some delay, the stored
       photon
      may be read
      out by sending in another control pulse; the frequency and structure of the recovered photon depends upon the
       shape
       of this readout pulse. The control pulse shapes which optimise the memory efficiency may be generalised for resonant quantum memory schemes \cite{gorshkov}.

       The results in this paper are one dimensional, in the sense that they apply to an ensemble which is much longer,
        along the propagation direction, than it is wide (with a Fresnel number \cite{langevin2} of order unity or less).
         A fully three dimensional model, capable of investigating the memory evolution with non-colinear signal and
          control fields, as well as the limits imposed on memory fidelity by spontaneous emission and motional
           decoherence, is the subject of a parallel publication \cite{Karl}. In the one-dimensional limit,
            we show that the full propagation problem is simplified greatly by the use of an appropriate modal
        decomposition.

       In \cite{raymer2} it is shown
       that the
       Raman interaction of a CW driving field with an atomic vapour can be decomposed as a set of two-mode squeezing
       transformations between light and matter. These modes characterise both the temporal structure of the
        Stokes light produced,
       and the spatial structure of the correlated spin wave generated in the ensemble. Recent work \cite{Wojtek}
       has verified the exact nature of this squeezed mode decomposition, shedding light on
       the properties of such multimode interactions. In particular the authors show how canonical evolution of the field operators
       fixes the relationship between the input and output modes for the problem. This analysis of Stokes generation is closely connected to the description of the quantum memory we present here. We show in this paper that the structure of the dynamical equations describing both systems
        guarantees canonical
       evolution of the optical and spin-wave field operators. Their commutation relations are automatically preserved by
        the
       solution and we do not need to impose this property as an extra condition. That is, the mode structure is intrinsic to the classical electrodynamics of the problem, rather than being intrinsically quantum mechanical. We also show that, if group velocity
        dispersion can be neglected, the evolution of the quantum memory subject to a control pulse of arbitrary shape is related
         to the simple case
       of a CW control field by a coordinate transformation. The evolution has a natural decomposition in terms of a
       set of mutually orthogonal input and output modes, and these modes are found to be related by
       a time-reversal-like symmetry. Each mode is associated with a singular value which quantifies the efficiency of
        the
       transfer from the input light-field mode to the stored spin-wave mode. The magnitudes of the singular values
         depend only upon a single parameter which
       represents the strength of the coupling of the quantum memory to the incident signal photon.

\section{Model}
   The control pulse, together with the ensemble of three-level atoms, may be treated as a two-level system, whose
    absorptive properties are determined by the control field. In \cite{Raymer1} a fully quantum mechanical treatment
     for the spontaneous initiation of Raman scattering in a one-dimensional atomic ensemble is introduced. The arguments
      may be adapted to describe the propagation of a signal photon through the memory cell.

      The $i^{\mathrm{th}}$ atom in the ensemble with energy eigenstates $\ket{j_i}$, at position $z_i$ and time $t$, couples to the
       electromagnetic field via the
       dipole interaction Hamiltonian (setting $\hbar=1$):
   \begin{eqnarray}
\label{Hamraman} \mathcal{H}^i_{\mathrm{int}}= &- \sum_{j,k} \sigma^i_{jk}\mathsf{d}^i_{jk}.\mathsf{E}(t,z_i).
\end{eqnarray}
The transition projection
      operators are defined by
   \begin{equation}
   \label{definesigma}
   \sigma^i_{jk}(t=0) \equiv \out{j_i}{k_i},
   \end{equation} in the Heisenberg picture. The electric field $\mathsf{E}(t,z)=\mathsf{E}_s(t,z) + \mathsf{E}_c(t,z)$ is made up of
    a classical
control field with central frequency $\omega_c$ and polarisation unit vector $\epsilon_c$,
\begin{equation}
\label{controlfield} \mathsf{E}_c(t,z)=\epsilon_c E_c \mathcal{E}(t,z) \exp{[-\mi \omega_c(t-z/c)]} +
 \mathrm{c.c.},
\end{equation}
and a single photon signal field with polarisation unit vector $\epsilon_s$, written in quantized notation as
\begin{equation}
\label{signalfield} \mathsf{E}_s(t,z)=\epsilon_s \mi \int_0^\infty g(\omega) a(t,\omega) \exp[{\mi \omega z/c}] \,
\rmd\omega + \mathrm{h.c.}.
\end{equation}
The spatio-temporal dependence of the control field is contained in the dimensionless envelope function
 $\mathcal{E}(t,z)$, which is scaled so that its maximum value is $1$. $E_c$ is then the peak amplitude of the
  control field.
In the Heisenberg picture, the mode annihilation operators satisfy the usual equal-time
    commutation relations
   \begin{equation}
   \label{acommutator}
   \left[a(t,\omega),a^\dagger(t,\omega')\right]=\delta(\omega-\omega').
   \end{equation}
The mode amplitude is given by $g(\omega)\equiv \sqrt{\frac{ \omega}{4 \pi \epsilon_0 \mathcal{A} c}}$, where
 $\mathcal{A}$ is the cross-sectional area of the region within the ensemble excited by the optical fields, $c$
  is the speed of light and $\epsilon_0$ is the
  permittivity of free space \cite{loudon}. We have also
defined the transition dipole moments by
\begin{equation}
\label{dipole} \mathsf{d}^i_{jk} \equiv \bra{j_i}\mathsf{d}^i\ket{k_i},
\end{equation}
where $\mathsf{d}^i$ is the electric dipole moment operator for the
$i^{\mathrm{th}}$ atom.

       The signal pulse amplitude is represented by the `slowly varying' operator
   \begin{equation}
   \label{defineA}
   A(t,z) \equiv \exp{[\mi \omega_s(t-z/v_s)]} \frac{1}{\sqrt{2\pi}}\int_{-\infty}^{\infty}
    a(t,\omega)\exp{[\mi \omega z/v_s]}\,\rmd\omega,
   \end{equation}
   where $\omega_s$ is the centre frequency of the pulse, and $v_s$ is the group velocity of the signal field.
   The spin wave to which the signal excitation is transferred is represented by the corresponding operator \cite{raymer2}
   \begin{equation}
   \label{defineB}
   B(t,z) \equiv \frac{1}{\delta z \sqrt{n\mathcal{A}}}\sum_\beta \sigma^\beta_{13}(t)\exp{\left\{-\mi
   \left[(\omega_s-\omega_c)
   t-\left(\frac{\omega_s}{v_s}-\frac{\omega_c}{v_c}\right)z_\beta\right]\right\}},
   \end{equation}
   with $v_c$ the group velocity of the control field.
   Here the index $\beta$ runs over all the atoms lying within a thin slice of the ensemble of thickness
    $\delta z$, centred at position $z$. $n$ is the atomic number density.
   The operators $A$ and $B$ act as continuous-mode bosonic annihilation operators for the signal pulse and atomic
    ensemble respectively; they satisfy the canonical commutation relations
   \begin{eqnarray}
   \label{commutatorA}
   [A(t,0),A^\dagger(t',0)]=&\delta(t-t'),\\
   \label{commutatorB}
   [B(0,z),B^\dagger(0,z')]=&\delta(z-z'),
   \end{eqnarray}
   where the Dirac delta function $\delta(x)$ is the derivative of the Heaviside step function
   \begin{equation}
\label{Heavy}
\Theta(x)=\cases{1&for $x \geq 0$;\\
0&for $x<0$.\\}
\end{equation}

\section{Propagation}
   Provided that the common detuning $\Delta$ of the signal and control pulses from single photon resonance is much
   larger than
    the signal photon bandwidth $\delta$, the excited state $\ket{m_i}$ can be adiabatically eliminated.
     If
     the ensemble is prepared in the collective groundstate $\ket{0}\equiv \bigotimes_i \ket{1_i}$, and if
      the
      population of the metastable state $\ket{3_i}$ is assumed to remain negligible, a linear theory can be used.
       The Maxwell-Bloch
       equations, in the slowly
       varying envelope approximation, are then found to be:
   \begin{equation}
   \label{Maxwell-Bloch1}
   \partial_t B(t,z) = -\gamma^*\mathcal{E}^*(t,z)A(t,z),
   \end{equation}
   and
   \begin{equation}
   \label{Maxwell-Bloch2}
    \left[\partial_z + \frac{1}{v_s}\partial_t\right] A(t,z) = \gamma \mathcal{E}(t,z)B(t,z).
   \end{equation}
   The coupling constant $\gamma$ is given, to first order, by
   \begin{equation}
   \label{definegamma}
   \gamma = \sqrt{\frac{n  \omega_s}{2\epsilon_0 c}}\frac{d_s^* d_c E_c}{\Delta},
   \end{equation}
   with $d_s$, $d_c$ the dominant transition dipole moments:
   \begin{equation}
   \label{dsdc}
   d_s\equiv \mathsf{d}^i_{m1}\cdot\mathsf{\epsilon}_s\; ;\qquad d_c \equiv \mathsf{d}^i_{m3}\cdot\mathsf{\epsilon}_c\,,\qquad
    \forall i.
   \end{equation}
   The dephasing of the material excitation $B$ can be modelled by appending a decay term and a Langevin operator
   \cite{langevin1} to
    (\ref{Maxwell-Bloch1}), but here for simplicity we neglect the effects of decoherence\footnote{We have also
     neglected
     the Stark shift of the levels $\ket{1_i}$ and $\ket{3_i}$. The former, induced by the presence of a
     single photon,
      is truly negligible; the latter is larger, but the control pulse centre frequency can always be chirped
       appropriately to cancel its
       effect.}: this remains a good approximation if the transverse coherence decay time of the spin wave is much
        larger than the
        duration of the signal pulse. We define the \emph{local time} $\tau \equiv t-z/v_s$, whence we obtain
   \begin{eqnarray}
   \label{MBs2.1}
   \partial_\tau B(\tau,z)=&-\gamma^* \mathcal{E}^*(\tau-\kappa z)A(\tau,z);\\
   \label{MBs2.2}
   \partial_z A(\tau,z)=&\quad\gamma \mathcal{E}(\tau-\kappa z)B(\tau,z).
   \end{eqnarray}
   Here we have expressed the functional form of the control pulse in terms of the \emph{dispersivity} $\kappa$,
    given by
   \begin{equation}
   \label{barv}
   \kappa \equiv \frac{1}{v_c} - \frac{1}{v_s}.
   \end{equation}
    In the case that dispersion can be neglected, so that
    $v_c\approx v_s$, the control field becomes a function of $\tau$ \emph{only}, and a Laplace transform over the
     spatial variable $z$ yields an analytic solution to the coupled equations (\ref{MBs2.1},\ref{MBs2.2})
      \cite{Raymer1}. This is the situation with which the bulk of this paper is concerned. However, even if
       $v_c \neq v_s$, the solutions can be obtained numerically \cite{Wojtek}. In the general case, for arbitrary $\kappa$, they have the
        following form:
   \begin{eqnarray}
   \label{generalsol1}
   A(\tau,L)=&\int_0^T C_A(\tau,\tau')A(\tau',0)\,\rmd \tau' + \int_0^L S_A(\tau,z')B(0,z')\,\rmd z',\\
   \label{generalsol2}
   B(T,z) =&\int_0^L C_B(z,z')B(0,z')\,\rmd z' - \int_0^T S_B(z,\tau')A(\tau',0)\,\rmd \tau',
   \end{eqnarray}
   where $L$ is the length of the ensemble, and $T$ is the duration of the readin process. The integral kernels $C_{A,B},S_{A,B}$ are Green's functions which propagate the boundary conditions $A(\tau,0)$ and $B(0,z)$. Because we write
    the solutions at the output of the ensemble, after the pulses have fully traveled through it, these
     are input-output,
     or scattering, relations.
\section{Unitarity and Mode Decomposition}
\subsection{Quantum memory dynamics}
   The coordinates $(\tau,z)$ may be discretized over an arbitrarily fine mesh, and then the above equations (\ref{generalsol1},\ref{generalsol2}) are
    represented (as in \cite{Wojtek}) to any chosen accuracy by the matrix equations
\begin{eqnarray}
\label{generalmatrix1}
\Av_L =& \Cv_A \Av_0 + \Sv_A \Bv_0;\\
\label{generalmatrix2} \Bv_T =& \Cv_B \Bv_0 - \Sv_B \Av_0,
\end{eqnarray}
with column vectors $\Av_0$,$\,\Bv_0$ replacing the initial signal and spin-wave amplitudes, and with $\Av_L$,$\Bv_T$ replacing the final signal and spin-wave
amplitudes. The integral kernels are replaced by the matrices $\Cv_{A,B}$ and $\Sv_{A,B}$. This
transformation should be unitary, since it describes the evolution of quantum mechanical operators, but it is
not
 immediately obvious how this is guaranteed by the equations (\ref{MBs2.1},\ref{MBs2.2}). As with an ordinary lossless
  beamsplitter, it is conservation of flux which constrains the solution. To see this, consider the following
  continuity relation
   implied by the evolution equations:
\begin{equation}
\label{flux1}
\partial_z A^\dagger(\tau,z)A(\tau,z) + \partial_\tau B^\dagger(\tau,z)B(\tau,z) = 0.
\end{equation}
Integrating this expression over a rectangle in $(\tau,z)$-space yields the expression
\begin{equation}
   \label{integrated}
   \int_0^\tau A^\dagger(\tau',z)A(\tau',z)\,\rmd \tau' +
   \int_0^z
   B^\dagger(\tau,z')B(\tau,z')\,\rmd z'=F_\tau(\tau)+F_z(z),
   \end{equation}
   where $F_z$ and $F_\tau$ are functions determined by the boundary conditions.
    Successively setting $(\tau,z)=(0,0),(0,L),(T,0),(T,L)$, we derive the
    flux-excitation conservation condition
   \begin{equation}
   \label{flux2}
   \Av_L^\dagger \cdot\Av_L + \Bv_T^\dagger \cdot \Bv_T = \Av_0^\dagger \cdot\Av_0 + \Bv_0^\dagger \cdot\Bv_0,
   \end{equation}
   which must hold for arbitrary initial amplitudes $\Av_0,\Bv_0$. Here the `dot' denotes the scalar product of two vectors: $\Av\cdot\Bv\equiv \Av^{\mathsf{T}}\Bv$, where the superscript $\mathsf{T}$ indicates matrix transposition. The dagger denotes Hermitian conjugation of \emph{operators}. When applied to an ordinary matrix, or a complex number, it is equivalent to complex conjugation. In order to avoid confusion, we do not use a dagger to indicate the composition of transposition with complex conjugation. We can cast the condition (\ref{flux2}) in a more compact form:
   \begin{equation}
   \label{Xform1}
   \Xv^\dagger\cdot\Xv =\Xv_0^\dagger\cdot\Xv_0,
   \end{equation}
   where we have defined
   \begin{equation}
   \label{Xdefs}
   \Xv\equiv\left(\begin{array}{c}\Av_L \\ \Bv_T \end{array}\right)=\Uv \Xv_0,
   \end{equation}
   with
   \begin{equation}
   \Xv_0 \equiv \left(\begin{array}{c}\Av_0 \\ \Bv_0 \end{array}\right)\; ;\quad \Uv \equiv \left(\begin{array}{cc}
   \Cv_A & \Sv_A \\ -\Sv_B & \Cv_B \end{array}\right)\;.
   \end{equation}
   (\ref{Xform1}) then tells us that the transformation $\Xv_0\rightarrow\Xv$ preserves the norm of $\Xv_0$, and this
    fixes $\Uv$ as unitary. The evolution of the operators describing the memory interaction is therefore canonical, as we would expect. This fact has important implications. In particular, multiplying out the relations $\Uv^{\mathsf{T}*}\Uv =\Uv\Uv^{\mathsf{T}*}=
    \Iv$ provides us with the conditions
    \begin{eqnarray}
    \label{blochmessiah1}
    \Cv_A^{\mathsf{T}*} \Cv_A + \Sv_B^{\mathsf{T}*}\Sv_B =& \Iv;\\
    \label{blochmessiah2}
    \Cv_B^{\mathsf{T}*} \Cv_B + \Sv_A^{\mathsf{T}*}\Sv_A =& \Iv,
    \end{eqnarray}
    along with a pair of antinormally ordered counterparts. We can now apply the Bloch-Messiah reduction
       \cite{braunstein} to (\ref{blochmessiah1}). For (\ref{blochmessiah1}), this consists in spectral decomposition \cite{nielsenchuang} of the positive
        Hermitian matrix products $\Cv_A^{\mathsf{T}*}\Cv_A$ and $\Sv_B^{\mathsf{T}*} \Sv_B$. These must commute with one another, and therefore they are both rendered diagonal in the same orthonormal basis. Similar
          conclusions follow for the remaining equations ((\ref{blochmessiah2}) and antinormal versions). As is shown in $\cite{Wojtek}$, the analysis reveals simple
           relationships between the singular value decompositions
    of the integral kernels, as follows:
   \begin{eqnarray}
   \label{CA}
   C_A(\tau,\tau')=&\sum_{i=1}^\infty \phi_i(\tau) \mu_i \psi_i(\tau'),\\
   \label{SB}
   S_B(\tau',z)=&\sum_{i=1}^\infty \phi_i(z) \lambda_i \psi_i(\tau'),\\
   \label{CB}
   C_B(z,z')=&\sum_{i=1}^\infty \varphi_i(z) \mu_i \chi_i(z'),\\
   \label{SA}
   S_A(\tau,z')=&\sum_{i=1}^\infty \varphi_i(\tau) \lambda_i \chi_i(z'),
   \end{eqnarray}
   where the functions $\{\phi_i\}$, $\{\psi_i\}$, $\{\varphi_i\}$, $\{\chi_i\}$,  each form a complete orthonormal
    basis, and where $\lambda_i$,$\mu_i$ are real, positive singular values for which
   \begin{equation}\label{pythagoras}\lambda_i^2 + \mu_i^2 = 1\quad \forall i.\end{equation}
   Equations (\ref{CA}-\ref{pythagoras}), when substituted into (\ref{generalsol1},\ref{generalsol2}), imply a
   set of independent beam-splitter transformations of ensemble modes and light-field modes.
\subsection{Application to Stokes scattering}
   Finally, we note that the above arguments apply in slightly altered form to the analysis of stimulated Stokes
    scattering presented in \cite{Wojtek}. In that case, Stokes photons and spin wave excitations are generated in pairs,
     and accordingly it is the flux \emph{difference} that is conserved by the Maxwell-Bloch equations (where now the
      operator $A$ represents the amplitude of the Stokes field):
   \begin{eqnarray}
   \label{StokesMaxwell1}
   \partial_\tau B(\tau,z)=&\gamma^* \mathcal{E}^*(\tau-\kappa z)A^\dagger(\tau,z);\\
   \label{StokesMaxwell2}
   \partial_z A(\tau,z)=&\gamma \mathcal{E}(\tau-\kappa z)B^\dagger(\tau,z).
   \end{eqnarray}
   The resulting flux-excitation conditions are altered slightly, corresponding to swapping the plus sign appearing in (\ref{flux2})
    for a minus sign. We can write them in the form
   \begin{eqnarray}
   \label{Stokesflux2}
   \Xv^{\dagger \mathsf{T}}\Zv\Xv = \Xv_0^{\dagger \mathsf{T}}\Zv \Xv_0;\\
   \label{Stokesantinormal}
   \Xv^\mathsf{T}\Zv\Xv^{\dagger} = \Xv_0^\mathsf{T}\Zv \Xv_0^{\dagger},
   \end{eqnarray}
   where
   $$
   \Zv \equiv \left(\begin{array}{cc}\Iv & 0 \\ 0 & -\Iv \end{array}\right)
   $$
   is the $z$-projection Pauli spin matrix.
   The solution to (\ref{StokesMaxwell1},\ref{StokesMaxwell2}) is now written
   \begin{equation}
   \label{Xform2}
   \Xv = \Cv \Xv_0 + \Sv \Xv_0^\dagger,
   \end{equation}
   where the matrices $\Cv$,$\Sv$ are of the form
   \begin{equation}
   \label{CS}
   \Cv \equiv \left(\begin{array}{cc} \Cv_A & 0 \\ 0 & \Cv_B \end{array}\right)\;;\quad \Sv \equiv \left(\begin{array}
   {cc}
    0 & \Sv_A \\ -\Sv_B & 0 \end{array}\right).\end{equation}
Substitution of this ansatz into the sum of (\ref{Stokesflux2}) and (\ref{Stokesantinormal}) provides us with
the conditions
\begin{eqnarray}
\label{diag1}
\Cv^{\mathsf{T}*}\Zv \Cv + \Sv^{\mathsf{T}}\Zv \Sv^* =& \Iv;\\
\label{cross1} \Cv^{\mathsf{T}*}\Zv \Sv + \Sv^{\mathsf{T}} \Zv \Cv^* =& 0.
\end{eqnarray}
These relations fix the inverse transformation to (\ref{Xform2}) as
\begin{equation}
\label{inverseX} \Xv_0 = \Zv \Cv^{\mathsf{T}*} \Xv - \Zv \Sv^{\mathsf{T}}\Xv^\dagger,
\end{equation}
and substitution of this into the flux conditions (\ref{Stokesflux2},\ref{Stokesantinormal}) yields the
antinormal condition
\begin{equation}
\label{finalantinormal} \Cv\Cv^{\mathsf{T}*} - \Sv\Sv^{\mathsf{T}*} = \Zv.
\end{equation}
Inspection of the components of the equations (\ref{diag1},\ref{finalantinormal}) reveals all the conditions
required of the matrices $\Cv_{A,B},\Sv_{A,B}$ to apply the Bloch-Messiah reduction. This then yields a mode
decomposition for the Stokes scattering problem, which is given explicitly in \cite{Wojtek} and used to identify
the dynamics with that of the well-known two-mode squeezer. The property of separability which allows the dynamics to be understood in terms of independent modes is therefore also related to a form of flux conservation.

    We now show that when dispersion can be neglected, the relations (\ref{CA}--\ref{SA}), describing the memory, simplify even further. Under this approximation, only one set of
     functions $\{\phi_i\}$, which may be computed by numerical diagonalisation of a known function, is involved in the
      dynamics. This approximation is consistent with the requirement of adiabaticity $\delta \ll \Delta$ for a broadband
       signal photon,
       since the refractive index of the ensemble varies slowly with frequency far from resonance.

   \section{Dispersionless case}
    We return to the equations (\ref{MBs2.1},\ref{MBs2.2}), and set $\kappa=0$, so that the $z$-dependence of the control
     field $\mathcal{E}$ vanishes. We  introduce the
         dimensionless `control pulse area'
   \begin{equation}
   \label{pulsearea}
   \epsilon=\epsilon(\tau)\equiv \gamma \sqrt{\frac{L}{E}}\int_0^\tau |\mathcal{E}(\tau')|^2 \,\rmd \tau',
   \end{equation}
   where the factor $E \equiv \int_0^T |\mathcal{E}(\tau')|^2\,\rmd \tau'$ appears in order to normalise
    this new variable so that it takes values between
     $0$ and
     $C\equiv\gamma \sqrt{LE}$.
   We also introduce a normalised spatial variable $\zeta \equiv \gamma z \sqrt{E/L}$. Again, $\zeta$ runs from $0$ to
    $C$. For simplicity we have set $\gamma$ to be real; alternatively any complex phase in $\gamma$ can
     always be incorporated into the field $\mathcal{E}$.
   In this new coordinate system the equations (\ref{MBs2.1},\ref{MBs2.2}) become
   \begin{equation}
   \label{newsystem}
   \partial_\zeta \alpha = \beta\; ;\;\;\; \partial_\epsilon \beta = -\alpha,
   \end{equation}
   where we have defined the dimensionless annihilation operators $\alpha(\epsilon,\zeta) \equiv \sqrt{E}A(\tau,z)/
   \mathcal{E}(\tau)$, and $\beta(\epsilon,\zeta)\equiv \sqrt{L}B(\tau,z)$. We have now eliminated the variation of the
    control field from the dynamical equations; the problem is reduced to that of a CW control field coupled to the
     ensemble for a brief time.
   The solutions for the signal amplitude $\alpha_C(\epsilon)\equiv \alpha(\epsilon,C)$ at the exit face of the
    ensemble,
    and the spin wave amplitude
$\beta_C(\zeta)\equiv\beta(C,\zeta)$ at the end of the readin process, are given in terms of the
initial conditions $\alpha(\epsilon,\zeta=0)=\alpha_0(\epsilon)$
     and $\beta(\epsilon=0,\zeta)=\beta_0(\zeta)$ by the expressions
\begin{eqnarray}
\label{exitfacea}
 \alpha_C(\epsilon)=&\int_0^C \left[G_1(\epsilon-x,C)\alpha_0(x)
+G_0(C-x,\epsilon)\beta_0(x)\right]\,\rmd x,\\
 \label{exitfaceb}
 \beta_C(\zeta)=&\int_0^C \left[G_1(\zeta-x,C)\beta_0(x)
-G_0(C-x,\zeta)\alpha_0(x)\right]\,\rmd x.
\end{eqnarray}
 There are now just two integral kernels, given by
   \begin{eqnarray}
   \label{greens1}
   G_0(p,q)\equiv& J_0 \left(2\sqrt{pq}\right),\\
   \label{greens2}
   G_1(p,q)\equiv& \delta(p) - \Theta(p)\sqrt{\frac{q}{p}}J_1\left(2\sqrt{pq}\right),
   \end{eqnarray}
   with the $n^{\mathrm{th}}$ Bessel function of the first kind denoted by $J_n$,
and where the Heaviside step function (\ref{Heavy})
 ensures that the convolutions in
(\ref{exitfacea},\ref{exitfaceb}) respect causality.
 We write the matrix representation of the solution (\ref{exitfacea},\ref{exitfaceb}) as
\begin{eqnarray}
\label{matrixexita}\alphav_C =& \Gv_1 \alphav_0 + \Gv_0 \betav_0\,;\\
\label{matrixexitb}\betav_C =& \Gv_1 \betav_0 - \Gv_0 \alphav_0.
\end{eqnarray}
Note that both of the matrices $\Gv_1$ and $\Gv_0$ are \emph{persymmetric}. That is, they are both symmetric
about their main anti-diagonal (or \emph{skew} diagonal). To see this, observe that the integral
 kernel $G_1$ (as it appears in (\ref{exitfacea})) varies only as a function of the quantity $\epsilon-x$.
  That is to say, its contours are all parallel to the line $\epsilon=x$, which corresponds to the main diagonal of
   $\Gv_1$. $\Gv_1$ is therefore invariant under reflection about its main anti-diagonal. The corresponding symmetry for
    $\Gv_0$ follows from the Hermiticity of the kernel $G_0(p,q)$ in its arguments. These symmetries allow us to
     decompose the kernels using
input and output modes related by time reversal (or equivalently space reversal):
   \begin{eqnarray}
   \label{simpleform0}
   G_0(C-x,\epsilon) =& \sum_{i=1}^\infty \phi_i(\epsilon)\lambda_i \phi_i(C-x),\\
   \label{simpleform1}
   G_1(\epsilon-x,C) =& \sum_{i=1}^\infty \phi_i(\epsilon)\mu_i \phi_i(C-x),
   \end{eqnarray}
   where there is now just a single set of real modefunctions $\{\phi_i\}$ such that
   \begin{equation}
   \label{orthonormal}
   \int_0^{C} \phi_i(x)\phi_j(x)\,\rmd x
   = \delta_{ij}.
   \end{equation}

\section{State transfer}
   Following the treatment in \cite{raymer2} we define output mode operators using the mode expansions
   \begin{eqnarray}
   \label{outputA}
   \alpha_C(\epsilon) \equiv& \sum_i \phi_i(\epsilon)A_i,\\
   \label{outputB}
   \beta_C(\zeta) \equiv& \sum_i \phi_i(\zeta) B_i.
   \end{eqnarray}
   For the input operators, we use the time-reversed expansions
   \begin{eqnarray}
   \label{inputA}
   \alpha_0(\epsilon) \equiv& \sum_i \phi_i(C-\epsilon)a_i,\\
   \label{inputB}
   \beta_0(\zeta) \equiv& \sum_i \phi_i(C-\zeta )b_i.
   \end{eqnarray}
   The solutions (\ref{exitfacea},\ref{exitfaceb}) can now be written in the form
   \begin{eqnarray}
   \label{bsmodesA}
   A_i =& \mu_i a_i + \lambda_i b_i\,,\\
   \label{bsmodesB}
   B_i =& \mu_i b_i - \lambda_i a_i\,,
   \end{eqnarray}
   with
   \begin{equation}
   \label{operatorscommutation}
   [A_i,A_j^\dagger]=[B_i,B_j^\dagger]=[a_i,a_j^\dagger]=[b_i,b_j^\dagger]=\delta_{ij}.
   \end{equation}

\begin{figure}
    \begin{center}
       \includegraphics[width=9cm]{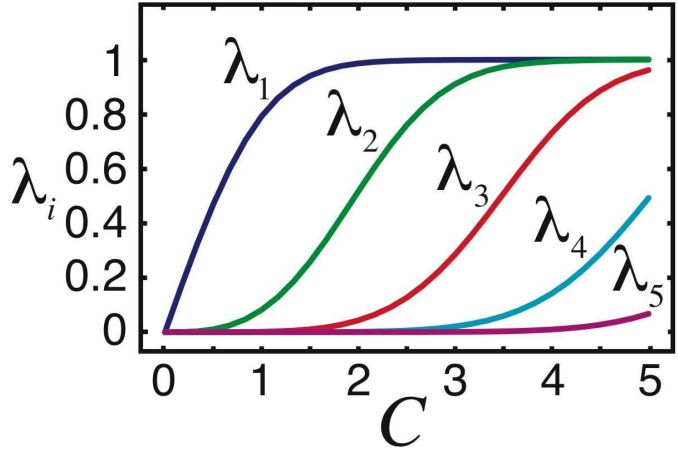}
       \caption{The five largest singular values of the kernel $\Gv_0$,
       plotted as a function of the coupling parameter $C$. The singular values are necessarily positive; for clarity
        they are plotted both above and below the origin.}
       \label{Figure2}
       \end{center}
       \end{figure}

   Equations (\ref{bsmodesA},\ref{bsmodesB}) show that the interaction of a signal photon with the memory cell can be
    viewed as a beamsplitter transformation
    on a mode-by-mode basis. Maximising the fidelity of the state transfer then amounts to minimising the
    terms $\mu_i a_i$.
   The magnitudes of the $\mu_i$ are determined by the coupling parameter $C$; a large value corresponds to an
    optically thick ensemble --- a high absorption. In figure \ref{Figure2} the singular values of the
     matrix $\Gv_0$ are plotted as a function of $C$. These are found by
     multiplying (\ref{simpleform0}) through by the $i^{\mathrm{th}}$ input mode and integrating, using (\ref{orthonormal}). We obtain the following relation:
     \begin{equation}\label{singvalsG0}
    \int_0^C G_0(C-y,x)\phi_i(C-y)\,\rmd y = \lambda_i \phi_i(x),\end{equation}
    corresponding to the simple eigenvalue equation
      \begin{equation}\label{eigenvals}
    \int_0^C J_0(2\sqrt{xy})\phi_i(y)\,\rmd y = \lambda_i \phi_i(x),\end{equation} which we solve numerically
    using a $500$ by $500$ square grid.
     We see that the interaction is dominated by the lowest mode for small $C$, but as $C$ increases, higher modes become
     significantly coupled. Modematching is therefore particularly important when the interaction is weak,
      as is the case for large detuning. Near-complete state transfer for the lowest mode ($\lambda_1\approx 1$)
       can be achieved for an ensemble
      with a coupling strength $C\geq 2$.
       \section{Modematching}
\begin{figure}
       \begin{center}
       \includegraphics[width=9cm]{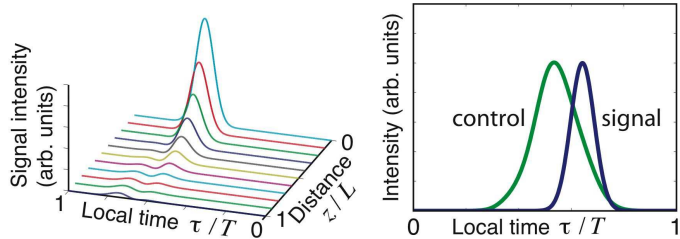}
       \caption{Top: the intensity $\expect{A^\dagger(\tau,z)A(\tau,z)}$ of a Gaussian signal photon as it propagates
        through an atomic ensemble with $C=2$. That is, a Fock state $\ket{1_\xi}\equiv a_\xi^\dagger\ket{\mathrm{vac}}$, where $a_\xi \equiv \int_{-\infty}^\infty \xi(\tau)A(\tau,0)
       \,\rmd \tau$ destroys a photon with a wave packet amplitude $\xi(\tau)\propto
        \exp{\{-2\ln 2[(\tau-\tau_0)/\sigma]^2\}}$,
        with $\sigma=1/8$, $\tau_0=T/2$ and $T=L=1$. $\ket{\mathrm{vac}}$ denotes the vacuum. The absorption is nearly
         complete; it is limited by the quality of the modematching optimisation. Bottom: the optimised
       control field intensity (blue/black) is shown alongside the initial signal field intensity (green/grey). The
        latter has
        been scaled for clarity. The lowest mode (corresponding to the largest singular value $\lambda_1$) was used
         in the optimisation.}
       \label{Figure3}
       \end{center}
       \end{figure}

   The efficiency of the memory is greatest for the lowest mode $\phi_1$. Therefore we require that the shape of
    this mode
    in real time matches the temporal structure of the signal photon wavepacket we send into
     our memory cell.
   The relationship between the shapes of the normalised real time input modes $\Phi_i(\tau)$, and the mode shapes as a function of
    control pulse area $\epsilon$, is
    determined by the temporal shape of the control field. The connection is straightforward,
    \begin{equation}
    \label{connection}
    \Phi_i(\tau) = \sqrt{\frac{C}{E}}\mathcal{E}(\tau)\phi_i[C-\epsilon(\tau)],
    \end{equation} but not analytically invertible, so a simple optimisation was performed in order to find the control
     pulse shape which matches a
     given input photon to the memory mode. The result for a signal photon with a Gaussian input profile, matched to the
      mode
         $\phi_1(C-\epsilon)$ with the highest
         coupling, and the simplest shape, is shown in figure \ref{Figure3}, for $C=2$. The photon is almost
         completely absorbed, although a small portion remains, due to the limitations of the modematching
         optimisation.

       \section{Readout}
       Once a properly modematched photon has been read in to the quantum memory, the ensemble is left in the output mode $\phi_1(\zeta)$, with probability amplitude $\lambda_1$. We now consider the effect of sending a second control
        pulse, propagating in the same direction as the initial control pulse, into the ensemble. The centre frequency, bandwidth and intensity of this
        \emph{readout} pulse may differ from that of the first control pulse (herein the \emph{readin} pulse). We can use
        all of the results developed thus far to analyse the interaction of the readout pulse with the ensemble,
        but we should be careful to keep track of any parameters which differ between the readin and readout
        stages. Let us use a superscript $r$ to indicate those quantities associated with the readout. If there
        is no significant change in the control and signal group velocities at readout (so that $v_c^r\approx
        v_s^r \approx v_c \approx v_s$), then the readout spin wave annihilation operator $B^r$ is phasematched to the
         readin spin wave operator $B$. That is, since the Stokes shift $\omega_s-\omega_c=\omega_s^r-\omega_c^r$
          is fixed,
         the phase factor
        appearing in (\ref{defineB}) is unchanged at readout, and we have $B^r(0,z)=B(T,z)$. Here again we have neglected any decoherence of the spin wave over the storage period. This provides us with
        one boundary condition; the second is that the signal field begins in its vacuum state at the start of
        the readout process: 
        \begin{equation}\label{boundaryA}
        \expect{A^{r\dagger}(\tau,0)A^r(\tau,0)}=0\, , \qquad \forall \tau.
        \end{equation}

\begin{figure}
       \begin{center}
       \includegraphics[width=9cm]{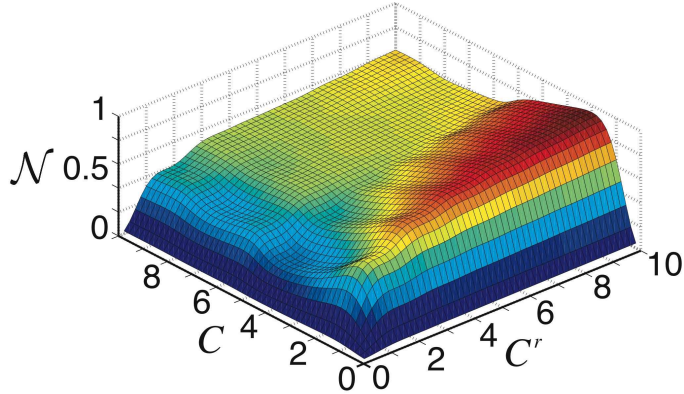}
       \caption{The photon retrieval probability $\mathcal{N}$ as a function of the readin and readout coupling parameters $C$ and $C^r$.}
       \label{Figure4}
       \end{center}
       \end{figure}

        At the end of the readout process, some portion of the stored excitation has been transferred back to the
        optical field. The efficiency of the readout depends upon the degree to which the spin wave mode
        $\phi_1(Cz/L)$ (written in terms of the ordinary spatial variable $z$)
        overlaps with the input modes for the readout process, which are of the form
        $\phi^r_i\left[C^r(1-z/L)\right]$. The functions $\{\phi_i^r\}$ solve the readout eigenvalue
        equation\begin{equation}
        \label{readouteigen}
        \int_0^{C^r}J_0(2\sqrt{xy})\phi^r_i(y)\,\rmd y = \lambda^r_i \phi^r_i(x).
        \end{equation}
        Each readout input mode $\phi^r_i\left[C^r(1-z/L)\right]$ is transferred to the optical field with
         amplitude $\lambda^r_i$, according to the relation (\ref{bsmodesA}), with these amplitudes set by the
         size of the readout coupling parameter $C^r$. A measure of the fidelity of the memory is the
         expectation value of the output photon number operator $\mathcal{N}$, which is just the probability of retrieving a photon from the ensemble at readout, given that a single modematched photon was sent in with the readin pulse. This retrieval probability is given by 
         \begin{eqnarray}
         \nonumber \mathcal{N}\equiv&\int_0^{T^r}\expect{A^{r\dagger}(\tau,L)A^r(\tau,L)}\,\rmd \tau\\
         \nonumber =& \frac{1}{C^r}\int_0^{C^r}\expect{\alpha^{r\dagger}(\epsilon^r,C^r)\alpha^r(\epsilon^r,C^r)}\,\rmd \epsilon^r \\
         \nonumber =& \frac{1}{C^r}\sum_{i=1}^\infty \expect{A_i^{r\dagger}A_i^\dagger}\\
         \nonumber =& \frac{1}{C^r}\sum_{i=1}^\infty \lambda_i^{r2}\expect{b_i^{r\dagger}b_i^r}\\
         \label{number} =& \lambda_1^2\sum_{i=1}^\infty \lambda_i^{r2}f_i^2,
         \end{eqnarray}
         where in the penultimate step we have used the transformation (\ref{bsmodesA}) and the boundary condition (\ref{boundaryA}) (setting $\expect{a_i^{r\dagger}a_i^r}=\expect{b_i^{r\dagger}a_i^r}=0$), and where we have defined the $i^{\mathrm{th}}$ \emph{overlap}
         \begin{equation}
         \label{overlap}
         f_i \equiv \frac{\sqrt{C^r C}}{L}\int_0^L \phi_i^r\left[C^r(1-z/L)\right]\phi_1(Cz/L)\,\rmd z.
         \end{equation}
         In figure \ref{Figure4} the variation of $\mathcal{N}$ is plotted as a function of the readin and readout coupling parameters $C$ and $C^r$.
       We observe that the retrieval probability is low if either $C$ or $C^r$ is small, as we expect from a weakly driven interaction. However, even if the photon is almost completely absorbed at readin, as is the case for the region $C\geq2$, the probability of recovering the stored photon approaches unity only slowly as $C^r$ is increased. Furthermore, as the readin process is driven harder (as $C$ rises above $2$), the retrieval probability falls. These
 observations are explained by considering the overlaps $f_i$ of the readout modes with the spin wave mode. For small $C$, $\phi_1(Cz/L)$ is monotonic, and relatively flat. It therefore has a large overlap $f_1$ with the lowest readout mode, which for moderate values of $C^r$ is essentially the mirror image of the spin wave mode. As $C$ increases, the spin wave mode becomes more asymmetric, so that its overlap with the lowest readout mode falls. It is then necessary to pump the readout process harder, so that higher readout modes, with which the spin wave overlaps significantly, are efficiently coupled to the optical field. The readout efficiency therefore grows as the readin coupling $C$ is decreased. But as $C$ becomes small, so also does the probability that the photon was stored during the readin process at all. The retrieval probability is maximised along the line $C\approx 2$, which represents the optimal coupling for the readin process. However, a readout coupling parameter in excess of $10$ is required to achieve $\mathcal{N}\geq 0.95$.
 
   The time reversal symmetry between the input and output modes makes the readout for this scheme a non-trivial problem. Simply repeating the readin process (so that $C=C^r$) results in poor performance of the memory.
   However, the dramatic increase in coupling strength required to extract the stored excitation fully, may make a na\"{i}ve increase in control pulse energy at readout prohibitively difficult to realise. An alternative method to boost the coupling is by significantly reducing the bandwidth of the readout pulse, along with its detuning $\Delta^r$. The photon recovered from the memory in this way would be frequency-shifted (according to the Raman resonance condition), and temporally stretched (since its bandwidth would be greatly diminished as well). The temporal profile of the emitted photon would then be determined by the shape of the readout pulse, as well as the shapes of the output modes involved in the readout process. A `photon transducer' with the ability to store broadband photons and convert them into narrowband photons with controllable frequency and shape could prove to be a useful tool for optical quantum information processing.
   
   Finally, we comment that the symmetry of the modes suggests the possibility of reading the stored excitation using a readout pulse propagating in the \emph{reverse} direction. Switching the propagation direction of the readout pulse would send $\phi_i(Cz/L)\rightarrow \phi_i\left[C(1-z/L)\right]$, so that the spin wave mode would overlap exactly with the lowest readout mode with $C=C^r$. Unfortunately, a counter propagating readout pulse would not have sufficient momentum to convert the spin wave into a counter propagating anti-Stokes photon. Another way to say this is that the readin and readout spin wave operators are not phasematched in this situation; the operator $B^r$ acquires
   a rapidly spatially varying phase factor, so that the overlap integrals $f_i$ vanish. However, an implementation of this scheme in the solid state (for example, replacing the atomic ensemble with an ensemble of semiconductor charge quantum dots \cite{sham}), might allow for the possibility of controlling the ensemble level structure with external fields. Then this kind of reverse readout might be achieved with the use of \emph{quasi}-phasematching \cite{boyd}, in which the sign of the readout coupling parameter $C^r$ is periodically flipped along the length of the ensemble. If the wavevector of this modulation matches the size of the wavevector mismatch between the readin and readout spin waves, the readout photon can be retrieved.  Alternatively, an adiabatic reversal of the sign of the Stokes shift over the whole ensemble, between the readin and readout stages (so that the energies of the states $\ket{3_i}$ and $\ket{1_i}$ are swapped at readout), would enable the phasematching of a high frequency counter propagating readout pulse, with the spin wave transferred to a counter propagating photon with lower frequency. Exploration of the possibilities for coherent manipulation of collective quanta in gaseous and solid-state ensembles will form the basis of continuing research.
   
   \section{Transverse profile}
   Here we briefly show that the modes $\{\phi_i\}$ also describe the transverse profile of the signal field. If the operators $\alpha$ and $\beta$ are allowed to depend upon the radial displacement $r$ from the $z$-axis, the Maxwell-Bloch equations, in the paraxial approximation \cite{raymer3}, take the modified form
   \begin{eqnarray}
   \label{MB3.1}
   \left[\frac{1}{4C}\nabla_\rho^2 + \mi \partial_\zeta \right] \alpha(\rv,\zeta,\epsilon)&=\mi \beta(\rv,\zeta,\epsilon)\,;\\
   \label{MB3.2}
   \partial_\epsilon \beta(\rv,\zeta,\epsilon)&= -\alpha(\rv,\zeta,\epsilon),
   \end{eqnarray}
   where $\rv$ is a normalised position vector for a point in the plane orthogonal to the propagation direction, a radial distance $r=\sqrt{2Lv_s/\omega_s} \rho$ from the $z$-axis. The extra term $\nabla_\rho^2$ is a $2$-dimensional Laplacian in the dimensionless cylindrical polar coordinates parameterising the point $\rv$. Here we have retained the assumption that the control field $\mathcal{E}$ has no radial dependence over the area $\mathcal{A}$. The dimensionless radial variable $\rho$ takes values from $0$ to $\rho_{\mathrm{max}}\equiv \sqrt{\omega_s/(2Lv_s)}r_{\mathrm{max}}$, where $\pi r_{\mathrm{max}}^2\equiv \mathcal{A}$. The Fresnel number $\mathcal{F}$ is given by \cite{langevin2}
   \begin{eqnarray}
   \nonumber \mathcal{F}&\equiv \left(\frac{\omega_s}{2\pi v_s}\right)\frac{\mathcal{A}}{L}\\ \label{fresnel}&=\rho_{max}^2.\end{eqnarray}
   The limit of the paraxial approximation is usually defined by $\mathcal{F}=1$, so we set $\rho_{\mathrm{max}}=1$.
   
   The solution to the equations (\ref{MB3.1},\ref{MB3.2}) is found by alternately Fourier transforming the variables $(\rv,\zeta)$ and $(\rv,\epsilon)$ \cite{raymer3} \cite{raymer3}. We find:
   \begin{eqnarray}
   \nonumber
   \alpha_C(\rv,\epsilon)&=\int_\mathcal{A}\,\rmd^2 \rv' \int_0^C\,\rmd x [R(\rv-\rv',C)G_1(\epsilon-x,C)\alpha_0(\rv',x)+\\ &\label{solR1} R(\rv-\rv',C-x)G_0(C-x,\epsilon)\beta_0(\rv',x)]\,;\\ \nonumber \beta_C(\rv,\zeta)&=\int_\mathcal{A}\,\rmd^2\rv' \int_0^C\,\rmd x [R(\rv-\rv',\zeta-x)G_1(\zeta-x,C)\beta_0(\rv',x)-\\ &\label{solR2} R(\rv-\rv',\zeta)G_0(C-x,\zeta)\alpha_0(\rv',x)]\,,\end{eqnarray}
   where $\int_{\mathcal{A}}\,\rmd^2\rv'$ denotes an integral over a circular patch with unit radius, and where the diffraction kernel is given, up to a normalisation constant $N$, by
   \begin{equation}
   \label{diffract}
   R(\rv,x)\equiv N \exp{\left(-\mi C |\rv|^2/x\right)}/x.
   \end{equation}
   This kernel is singular for vanishing $x$, since the paraxial approximation breaks down in the near-field. However, in the limit that the interaction is negligible close to the exit face of the ensemble, we can make the replacement $R(\rv,x)\rightarrow R(\rv,C)$. The diffraction kernel then factorises out of the $\epsilon$ and $\zeta$ integrals in (\ref{solR1},\ref{solR2}). We introduce the spectral decomposition
   \begin{equation}
   \label{Rdecomp}
   R(\rv-\rv',C)=\sum_{j=0}^\infty \varphi_j(\rv)\sigma_j \varphi_j(\rv'),
   \end{equation} where the $\sigma_j$ are real eigenvalues and $\{\varphi_j\}$ is a complete orthonormal set of \emph{paraxial modes}, satisfying
   \begin{equation}
   \label{paraxial}
   \int_\mathcal{A} \varphi_j(\rv)\varphi^*_k(\rv)\,\rmd^2\rv=\delta_{jk}.
   \end{equation}
    We assume that the spatial profile of the signal field depends only upon the radial coordinate $\rho$. Cylindrical symmetry of the boundary conditions then allows us to neglect modes with any dependence on azimuthal angle. These cylindrically symmetric modes are then given by 
   \begin{equation}
   \label{phis}
   \varphi_j(\rho)=e^{\mi \rho^2} \phi_j^1(\rho^2),
   \end{equation}
   where the superscript $1$ indicates that the function $\phi_j^1$ solves the eigenvalue equation (\ref{eigenvals}) with $C$ set equal to $1$. We define input and output mode operators, as in (\ref{outputA},\ref{outputB},\ref{inputA},\ref{inputB}), by the relations:
   \begin{eqnarray}
   \label{inputAR}
   \alpha_0(\rv,\epsilon)\equiv \sum_{(i,j)=0}^{\infty} \phi_i(C-\epsilon)\varphi_j(\rho)a_{ij}\,;\\
   \label{inputBR}
   \beta_0(\rv,\zeta)\equiv \sum_{(i,j)=0}^{\infty} \phi_i(C-\zeta) \varphi_j(\rho) b_{ij}\,,\end{eqnarray}
   and
   \begin{eqnarray}
   \label{outputAR}
   \alpha_C(\rv,\epsilon)\equiv \sum_{(i,j)=0}^{\infty} \phi_i(\epsilon)\varphi_j(\rho)A_{ij}\,;\\
   \label{outputBR}
   \beta_C(\rv,\zeta)\equiv \sum_{(i,j)=0}^{\infty} \phi_i(\zeta) \varphi_j(\rho) B_{ij}\,,\end{eqnarray}
   with
   \begin{equation}
   \label{operatorscommutationR}
   [A_{ij},A_{kl}^\dagger]=[B_{ij},B_{kl}^\dagger]=[a_{ij},a_{kl}^\dagger]=[b_{ij},b_{kl}^\dagger]=\delta_{ik}\delta_{jl}.
   \end{equation}
   We then obtain the beamsplitter transformations:
   \begin{eqnarray}
   \label{transformR1}
   A_{ij}=\sigma_j\left(\mu_i a_{ij} + \lambda_i b_{ij}\right)\,;\\
   \label{transformR2}
   B_{ij}=\sigma_j\left(\mu_i b_{ij} - \lambda_i a_{ij}\right)\,.
   \end{eqnarray}
   The lowest mode is dominantly coupled, with $\sigma_1 = 0.995$. The shape of this dominant spatial mode is well approximated by a Gaussian with beam waist $w_s=1.45$ (in the normalised units of the variable $\rho$). The control should be slightly less tightly focussed, with a waist $w_c \geq 3 w_s$, to justify the approximation that the control field may be treated as a train of plane waves. If such appropriately focussed Gaussian beams are used for the signal and control fields, only the lowest paraxial mode is involved in the interaction. The index $j$ may be dropped from the relations (\ref{transformR1},\ref{transformR2}), and setting $\sigma_1\approx 1$, we recover the transformations (\ref{bsmodesA},\ref{bsmodesB}), and the one-dimensional treatment is justified.
\section{Experiment}
       We are currently developing an implementation of this scheme in atomic thallium vapour. The gross structure of
        thallium has a $\Lambda$-type triplet, where the excited $7s ^2S_{1/2}$ state is coupled strongly to the ground
         state $6p ^2 P_{1/2}$ and the metastable state $6p ^2 P_{3/2}$ (with an electric-dipole-forbidden radiative
          lifetime of $0.16$ s).
       The splitting of these last two levels is very large (around $8000$ $\mathrm{cm}^{-1}$), so that the signal and
        control pulses remain spectrally distinct, even when their bandwidths are large. The more stringent limit on the
         bandwidth of the photon that can be stored comes from the adiabaticity condition: in order that no real
          population is transferred to the excited state, atomic collisions must occur at a negligible rate
           \cite{raymerfluorescence}, and
          the signal bandwidth $\delta$ should be
           much smaller than the detuning $\Delta$. The interaction cross-section scales as the inverse square of this
           detuning, so a compromise must be made between the bandwidth $\delta$ that is stored and the
            length $L$ of the
            ensemble. We hope to demonstrate the storage and retrieval of ultraviolet pulses ($\sim390$ nm) with
             $\delta \approx 10$ nm and pulse lengths of around $100$ fs.

\section{Summary}
        We have shown that it is possible to map a broadband photon wavepacket into an atomic spin wave using a Raman interaction. The relationship between the input state, the storage state and the output is clarified by the introduction of a mode decomposition. The modes are expressed in terms of a single universal set of eigenfunctions of a particular integral kernel. The form of these eigenfunctions depends only on the
           coupling strength of the memory interaction, and not upon the shapes of the pulses.
\ack
        The authors gratefully acknowledge the support of Hewlett-Packard and the EPSRC (UK) through the QIP IRC (GR/S82176/01). MGR thanks the QIP IRC for support during his stay in Oxford, and also acknowledges the US NSF (AMOP) for support. The research of DJ was supported in part by The Perimeter Institute for Theoretical Physics.

\bibliography{references}

\begin{thebibliography}{10}

\bibitem{distributed}
J.~I. Cirac, A.~K. Ekert, S.~F. Huelga, and C.~Macchiavello.
\newblock Distributed quantum computation over noisy channels.
\newblock {\em quant-ph/9803017}, 1999.

\bibitem{cryptography}
Nicolas Gisin, Grigoire Ribordy, Wolfgang Tittel, and Hugo Zbinden.
\newblock Quantum cryptography.
\newblock {\em quant-ph/0101098}, 2001.

\bibitem{repeater}
H.-J. Briegel, W.~D{\"{u}}r, J.~I. Cirac, and P.~Zoller.
\newblock Quantum repeaters: The role of imperfect local operations in quantum
  communication.
\newblock {\em Phys. Rev. Lett.}, 81:5932--5935, 1998.

\bibitem{grover1}
Grover~L. K.
\newblock From schrodinger's equation to quantum search algorithm.
\newblock {\em American Journal of Physics}, 69(7):769--777, 2001.

\bibitem{shor1}
Peter Shor.
\newblock Polynomial-time algorithms for prime factorization and discrete
  logarithms on a quantum computer.
\newblock {\em SIAM J. Sci. Statist. Comput.}, 26:1484, 1997.

\bibitem{polzikscheme}
A.~E. Kozhekin, K.~M{\o}lmer, and E.~Polzik.
\newblock Quantum memory for light.
\newblock {\em Phys. Rev. A}, 62:033809, 2000.

\bibitem{Raymer1}
M.~G. Raymer and J.~Mostowski.
\newblock Stimulated {R}aman scattering: Unified treatment of spontaneous
  initiation and spatial propagation.
\newblock {\em Phys. Rev. A}, 24(4):1980--1993, 1981.

\bibitem{raymer2}
M.~G. Raymer.
\newblock Quantum state entanglement and readout of collective atomic-ensemble
  modes and optical wave packets by stimulated {R}aman scattering.
\newblock {\em J. Mod. Opt.}, 51(12):1739--1759, 2004.

\bibitem{Wojtek}
Wojciech Wasilewski and M.G. Raymer.
\newblock Pairwise entanglement and readout of atomic-ensemble and optical
  wave-packet modes in traveling-wave {R}aman interactions.
\newblock {\em quant-ph/0512157. Submitted to Phys. Rev. A}, 2005.

\bibitem{longdistance}
L-M. Duan, M.~D. Lukin, J.~I. Cirac, and P.~Zoller.
\newblock Long-distance quantum communication with atomic ensembles and linear
  optics.
\newblock {\em Nature}, 414:413--418, 2001.

\bibitem{3Dpaper}
L-M. Duan, J.~I. Cirac, and P.~Zoller.
\newblock Three-dimensional theory for interaction between atomic ensembles and
  free-space light.
\newblock {\em Phys. Rev. A}, 66:023818, 2002.

\bibitem{EIT1}
M.~Fleischhauer and M.~D. Lukin.
\newblock Quantum memory for photons: Dark-state polaritons.
\newblock {\em Am. Phys. Soc.}, 65:022814, 2002.

\bibitem{EITreview2}
M.~Fleischhauer, A.~Immamoglu, and J.~Marangos.
\newblock Electromagnetically induced transparency: Optics in coherent media.
\newblock {\em Rev. Mod. Phys.}, 77:633, 2005.

\bibitem{gorshkov}
Alexey~V. Gorshkov, Axel Andr{\'{e}}, Michael Fleischhauer, Anders~S
  S{\o}rensen, and Mikhail~D. Lukin.
\newblock Optimal storage of photon states in optically dense atomic media.
\newblock {\em quant-ph/0604037}, 2006.

\bibitem{langevin2}
M.~G. Raymer and I.~A. Walmsley.
\newblock {\em Progress in {O}ptics}.
\newblock {E}lsevier {S}cience {P}ublishers B.V., xxviii edition, 1990.

\bibitem{Karl}
K.~Surmacz et~al.
\newblock Storing broadband photons in an atomic quantum memory.
\newblock {\em Unpublished}, 2006.

\bibitem{loudon}
Rodney Loudon.
\newblock {\em The Quantum Theory of Light}.
\newblock Oxford University Press, 2004.

\bibitem{langevin1}
H.~Haken.
\newblock Cooperative phenomena.
\newblock {\em Rev. Mod. Phys.}, 47(1):97--121, 1975.

\bibitem{braunstein}
Samuel~L. Braunstein.
\newblock Squeezing as an irreducible resource.
\newblock {\em quant-ph/9904002}, 1999.

\bibitem{nielsenchuang}
M.~A. Nielsen and I.~L. Chuang.
\newblock {\em Quantum Computation and Quantum Information}.
\newblock Cambridge University Press, 2004.

\bibitem{sham}
Wang Yao, Ren-Bao Liu, and L.~J. Sham.
\newblock Theory of control of the spin-photon interface for quantum networks.
\newblock {\em Phys. Rev. Lett.}, 95:030504, 2005.

\bibitem{boyd}
Robert~W. Boyd.
\newblock {\em Nonlinear Optics}.
\newblock Academic Press, 1992.

\bibitem{raymer3}
M.~G. Raymer and I.~A. Walmsley.
\newblock Quantum theory of spatial and temporal coherence properties of
  stimulated {R}aman scattering.
\newblock {\em Phys. Rev. A}, 32(1):332--344, 1985.

\bibitem{raymerfluorescence}
M.~G. Raymer and J.~L. Carlsten.
\newblock Simultaneous observations of stimulated {R}aman scattering and
  stimulated collision-induced fluorescence.
\newblock {\em Phys. Rev. Lett.}, 39:1326, 1977.

\end{thebibliography}
\end{document}